\documentclass[twocolumn,aip,nofootinbib,jcp,reprint,amsmath,amssymb]{revtex4-2}

\usepackage{natbib}
\usepackage{amsmath}
\usepackage{graphicx}
\usepackage[hidelinks]{hyperref}
\usepackage{dcolumn}
\usepackage{bm}
\usepackage{epsfig,color,xspace,multirow,xr,bbold}
\usepackage[all]{xy}
\usepackage{setspace}
\usepackage{url}
\usepackage{soul}
\usepackage[colorinlistoftodos]{todonotes}
\usepackage{longtable}
\usepackage{tabularx}
\usepackage{adjustbox}
\usepackage{threeparttable} 
\usepackage{natbib,hyperref}
\usepackage{float}
\usepackage{placeins}
\usepackage[utf8]{inputenc}
\usepackage[T1]{fontenc}
\usepackage{pifont}
\usepackage{booktabs}
\usepackage[table]{xcolor}

\definecolor{wf}{RGB}{90,90,90}

\definecolor{rsh}{RGB}{145,70,220}

\definecolor{comp}{RGB}{60,120,255}

\definecolor{dh}{RGB}{40,170,80}

\definecolor{hyb}{RGB}{245,210,40}

\definecolor{meta}{RGB}{245,140,30}

\definecolor{gga}{RGB}{220,50,45}

\definecolor{deepgreen}{RGB}{0,100,0}

\usepackage{xcolor}
\definecolor{deeppink}{RGB}{255,20,147}

\usepackage[normalem]{ulem}
\usepackage{xcolor}

\newcommand\notsotiny{\@setfontsize\notsotiny\@vipt\@viipt}
\usepackage{soul} 
\newcommand{\orcid}[1]{\href{https://orcid.org/#1}{\includegraphics[width=8pt]{orcid.png}}}

\begin{document}

\title{
Potential for helicity-mediated reverse intersystem crossing in 
boron--nitrogen-substituted polycyclic aromatic hydrocarbons 
with inverted singlet-triplet gaps
}
\date{\today}

\author{
Atreyee Majumdar, 
Raghunathan Ramakrishnan$^*$ 
}
\email{ramakrishnan@tifrh.res.in} 
\thanks{\\ $^*$ Corresponding author}  
\affiliation{Tata Institute of Fundamental Research, Hyderabad 500046, India}


\begin{abstract}
\noindent 
The prospect of exothermic reverse intersystem crossing (rISC) in organic chromophores with inverted singlet–triplet (INVEST) excited states has stimulated considerable interest in next-generation emitters. 
However, realizing efficient spin conversion remains challenging because energetic inversion does not necessarily imply the appreciable spin–orbit coupling (SOC) required for rISC.
Here, we investigate direct ${\rm S}_1-{\rm T}_1$ SOC in 85 INVEST candidates comprising 32 azaphenalenes, 7 non-alternant hydrocarbons, and 46 boron–nitrogen-substituted polycyclic aromatic hydrocarbons (BNPAHs), all previously predicted to exhibit negative singlet–triplet gaps using a correlated wavefunction method at reliably optimized molecular geometries. 
While the azaphenalenes and non-alternant hydrocarbons exhibit negligible direct SOC, 28 of the 46 BNPAHs display finite coupling. 
Analysis of molecular geometries and natural transition orbitals suggests that SOC variations depend on molecular topology, orbital orientation, and electronic-state symmetry rather than nonplanarity or excited-state character alone.
We estimate direct-rISC rates with a Marcus model, identifying selected BNPAHs with rates approaching $10^7\,{\rm s}^{-1}$. 
\end{abstract}

\maketitle

The experimental report by Aizawa \emph{et al.} in 2022\cite{aizawa2022delayed}, suggesting delayed fluorescence (DF) in organic chromophores with inverted singlet-triplet (INVEST) excited states, generated considerable interest in organic optoelectronics.\cite{li2022organic,li2022down,won2023inverted,liu2026largely}
The reported negative singlet-triplet gap, DF lifetime of  $\approx$0.2~$\mu$s, and organic light-emitting diodes with an external quantum efficiency of 17\% suggested a route to triplet harvesting without the uphill thermal activation characteristic of conventional
thermally activated DF.\cite{aizawa2022delayed}
Intersystem crossing (ISC) transfers population from singlet to triplet states, whereas reverse intersystem crossing (rISC) enables triplet-to-singlet conversion and can repopulate the emissive singlet state to produce DF. 
These prospects motivated the search for INVEST chromophores combining favorable excited-state energetics with efficient spin conversion.
\begin{figure}[!htpb]
    \centering
    \includegraphics[width=\linewidth]{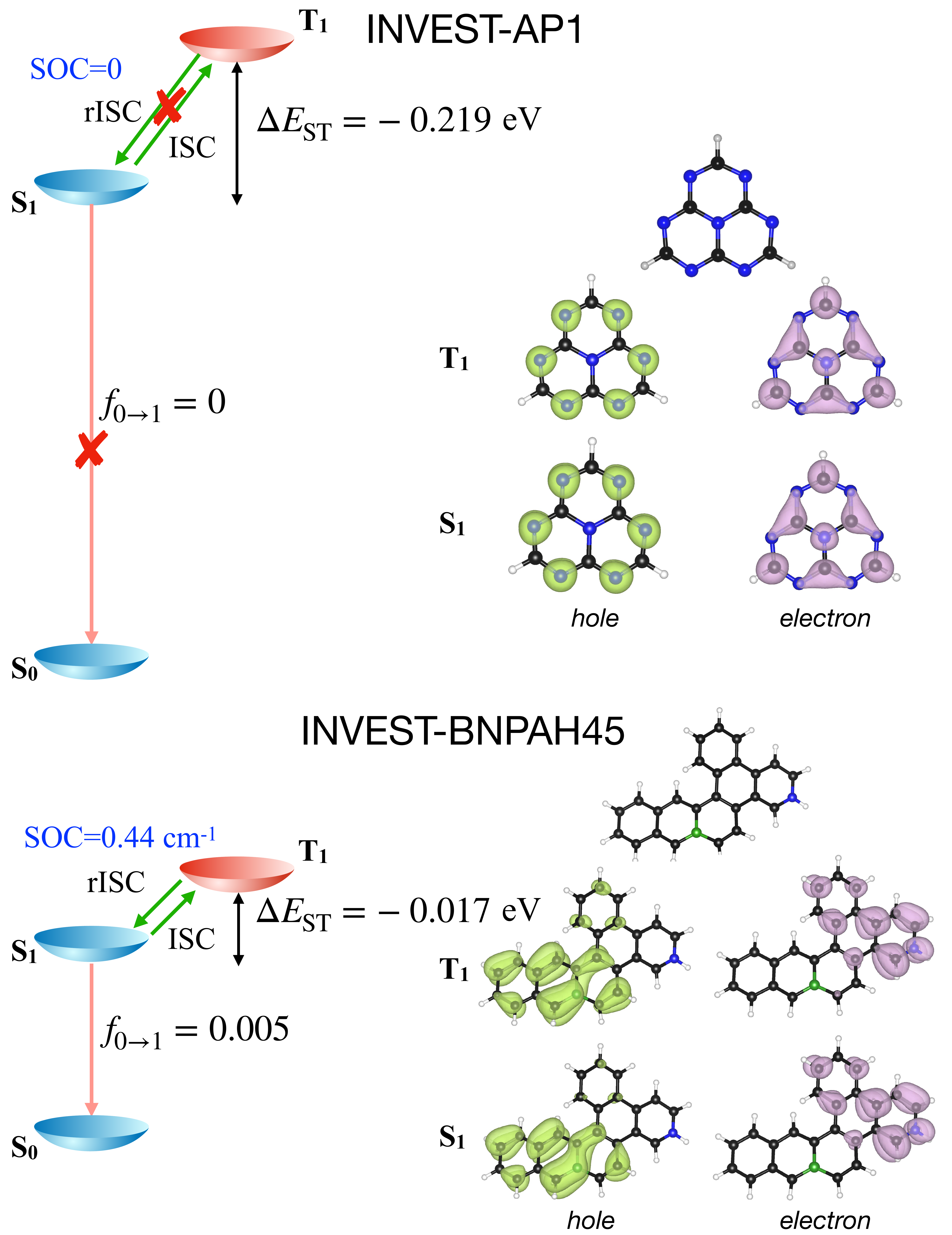}
    \caption{Schematic direct ${\rm T}_1\rightarrow {\rm S}_1$ pathway for heptazine (top) and BNPAH45 (bottom), along with the dominant hole and electron NTOs for ${\rm S}_1$ and ${\rm T}_1$. 
    }
    \label{fig:mechanism}
\end{figure}

A negative singlet-triplet gap, however, does not guarantee rapid rISC or efficient DF. A recent spectroscopic reinvestigation of the heptazine emitter HzTFEX$_2$ supported singlet-triplet inversion but found strongly suppressed ISC following direct ${\rm S}_1$ excitation and no detectable long-lived DF at room temperature.\cite{francener2026comprehensive} 
Higher-state-mediated rISC pathways have also been explored as possible mechanisms for facilitating spin conversion in INVEST systems.
Ricci \emph{et al.} discussed sequential ${\rm T}_1\rightarrow {\rm S}_2\rightarrow {\rm S}_1$ conversion, involving rISC to ${\rm S}_2$ followed by internal conversion to ${\rm S}_1$.\cite{ricci2025enhancing} 
Izu \emph{et al.} investigated spin-vibronic coupling in doped triangulenes, involving vibrational coupling between ${\rm T}_1$ and higher triplets (${\rm T}_n$) together with spin-orbit
coupling (SOC) between ${\rm T}_n$ and ${\rm S}_1$.\cite{izu2024reverse}
Both pathways may repopulate ${\rm S}_1$ and support DF even when direct ${\rm S}_1$-${\rm T}_1$ SOC is weak, although their efficiency depends on competing relaxation processes. These considerations
motivate the search for INVEST chromophores with appreciable direct ${\rm S}_1$-${\rm T}_1$ SOC.

The distinction between energetic inversion and direct ${\rm S}_1$-${\rm T}_1$  SOC is illustrated schematically in Fig.~\ref{fig:mechanism}, which compares heptazine with \textbf{BNPAH45}.
Although both molecules exhibit inverted singlet--triplet gaps at the correlated L-CC2 level, their direct ${\rm S}_1$-${\rm T}_1$  SOC magnitudes differ markedly. 
While direct SOC is negligible for heptazine at the level of theory employed, \textbf{BNPAH45} exhibits an appreciable coupling of  0.44~cm$^{-1}$.
Interestingly, the dominant ${\rm S}_1$ and ${\rm T}_1$ natural transition orbitals (NTOs) exhibit similar hole and electron distributions within each molecule. 
This contrast raises the question of which structural and electronic features enable appreciable direct singlet-triplet coupling in INVEST chromophores
composed predominantly of light elements.

For a direct SOC-mediated pathway, El-Sayed's rule suggests that coupling is generally weak between states of similar orbital excitation character, particularly when molecular symmetry imposes additional restrictions \cite{el1963spin}. Such limitations are important in chromophores composed of light elements, for which SOC is relatively small. 
Molecular nonplanarity offers a possible route to relaxing these restrictions. 
Schmidt \emph{et al.} showed that SOC and ISC in monoaza[5]helicenes are sensitive to molecular nonplanarity.\cite{schmidt2007intersystem} 
Bergwinkl \emph{et al.} subsequently found that helical distortion and the sulfur heavy-atom effect cooperate in enhancing SOC in a thiohelicene: a calculated coupling of 0.0035~cm$^{-1}$ in planar thiohexacene increased to 4.99~cm$^{-1}$ in the helical sulfur-containing system.\cite{bergwinkl2024enhanced} 
The enhancement in that study also involved higher triplet states; it therefore supports a structural route to SOC modulation rather than establishing direct rISC in INVEST systems. Yanagi \emph{et al.} further showed that heteroatom-containing helicenes can retain substantial configurational stability.\cite{yanagi2021asymmetric}

\begin{figure}[!htpb]
    \centering
    \includegraphics[width=\linewidth]{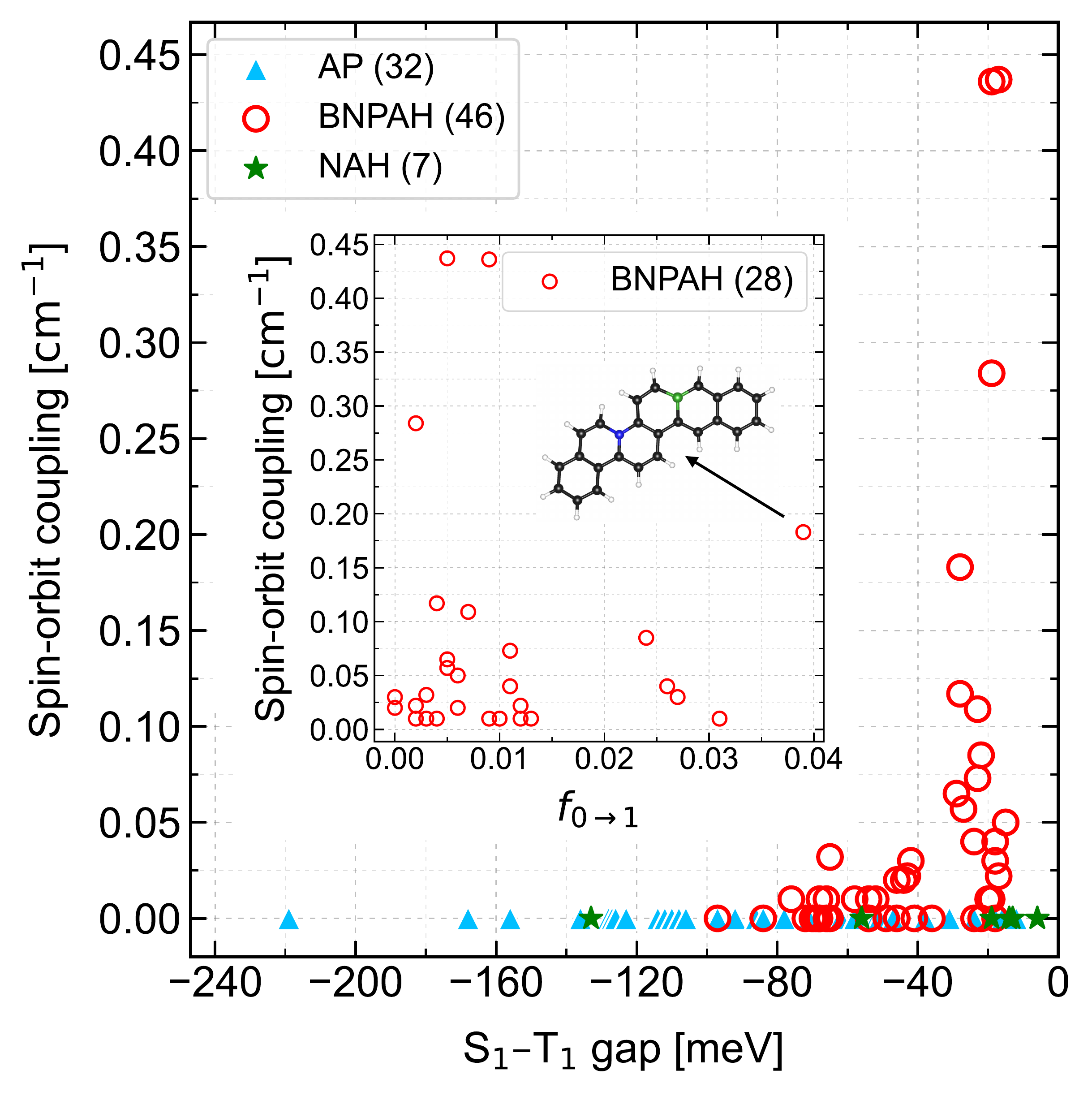}
    \caption{
    TDDFT Direct ${\rm S}_1$-${\rm T}_1$ SOC magnitude versus the L-CC2 vertical singlet-triplet gap for 85 INVEST chromophores. The
    inset shows oscillator strength ($f$) versus SOC magnitude for the
    28 BNPAHs exhibiting finite direct SOC.
    Colors distinguish azaphenalenes (blue), non-alternant hydrocarbons (green), and BNPAHs (red). 
}
    \label{fig:gap_soc}
\end{figure}

In an earlier high-throughput study, 30\,797 BN-substituted PAHs derived from 77 parent frameworks were screened, identifying 46 promising DFIST-BNPAHs, including BN-helicenes exhibiting through-space charge-transfer character.\cite{majumdar2025unlocking} 
The present investigation builds on those candidates by asking whether favorable inverted gaps are accompanied by direct ${\rm S}_1$-${\rm T}_1$ SOC and, if so, how structural distortion influences the coupling. 
We compare 46 BNPAHs with 32 azaphenalenes and 7 non-alternant hydrocarbons and estimate direct-rISC rates under a common kinetic model. 
Structures are shown in Figures~S1--S3 of the Supporting Information (SI). 
The focus is on the direct pathway rather than a complete treatment of higher-state-mediated spin conversion.

The ground-state geometries and vertical excitation
energies of the 85 INVEST candidates were taken
from our previous work.\cite{majumdar2026optimized}
The geometries were optimized at the
$\omega$B97M-D4/def2-TZVP level, and the lowest
singlet and triplet excitation energies were
evaluated using L-CC2/aug-cc-pVDZ.
Previous assessments\cite{majumdar2026insights} have shown that L-CC2 provides reliable estimates of inverted singlet--triplet gaps when benchmarked against theoretical best estimates\cite{loos2023heptazine,loos2025correction}.

We define the vertical singlet--triplet gap by 
\begin{equation}
\Delta E_{\mathrm{ST}}=E({\rm S}_1)-E({\rm T}_1),
\end{equation}
with $\Delta E_{\mathrm{ST}}<0$ denoting the inverted ordering. 
Direct ${\rm S}_1$--${\rm T}_1$ SOC calculations
were performed at the fixed ground-state
geometries using Tamm--Dancoff-approximated (TDA)
time-dependent density functional theory (TDDFT)
at the $\omega$B97M-D4/def2-TZVP level,
as implemented in ORCA.
The direct SOC magnitude is defined as the
norm of the three triplet-component couplings,
\begin{equation}
|V^{d}| =
\sqrt{\sum_{m=0,\pm1}
\left|
\langle {\rm S}_1 |
\hat H_{\mathrm{SO}} |
{\rm T}_{1,m} \rangle
\right|^2},
\end{equation}
where $m$ labels the three triplet spin sublevels.

\begin{figure*}[!htpb]
    \centering
    \includegraphics[width=\linewidth]{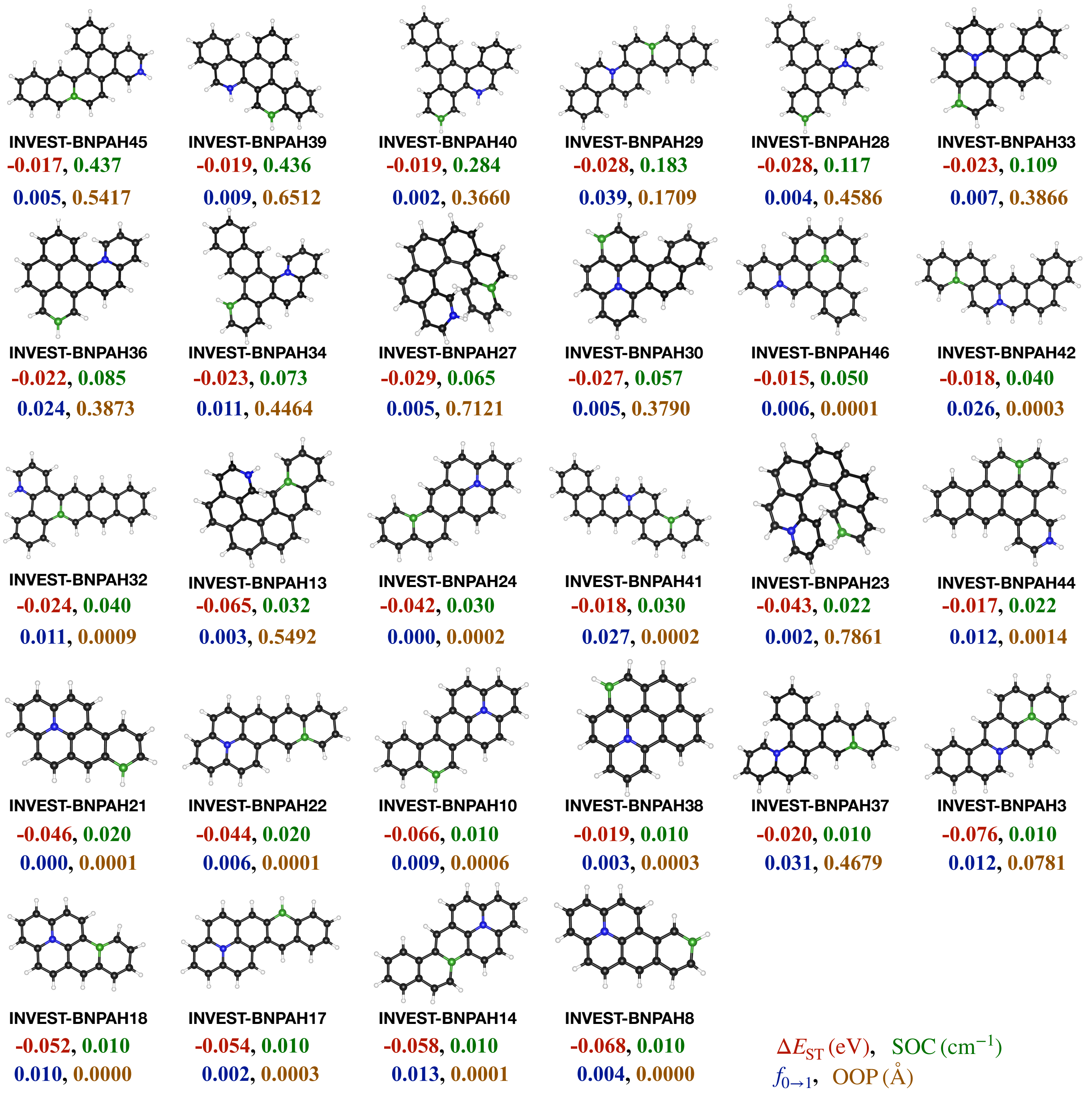}
    \caption{Molecular structures of the 28 INVEST-BNPAHs with finite direct ${\rm S}_1$-${\rm T}_1$ SOC. The L-CC2 singlet--triplet gap (eV) is shown in red, the TDDFT SOC norm (cm$^{-1}$) in green, the ${\rm S}_1$ oscillator strength (dimensionless) in blue, and the RMS out-of-plane displacement of non-hydrogen atoms (\AA) in brown. 
    }
    \label{fig:28BNPAHs}
\end{figure*}

Figure~\ref{fig:gap_soc} plots the direct ${\rm S}_1$-${\rm T}_1$ SOC magnitude against the L-CC2/aug-cc-pVDZ gap for all 85 INVEST molecules. 
All 32 azaphenalenes and 7 non-alternant hydrocarbons exhibit negligible couplings.
In contrast, 28 of the 46 BNPAHs have finite reported coupling, with the largest values reaching several tenths of a wavenumber. 
These contrasting trends highlight the distinct electronic and structural factors governing singlet-triplet inversion and direct SOC. 
The sensitivity of the calculated SOC magnitudes
to the choice of exchange--correlation functional
was examined using B3LYP, M06-2X, LC-BLYP, and
$\omega$B97M-D4 (Figure~S4 of the SI). 
The two range-separated functionals, $\omega$B97M-D4 and LC-BLYP, show closely correlated SOC magnitudes across the BNPAH set ($R^2=0.944$); in contrast, B3LYP gives considerably less consistent magnitudes ($R^2=0.123$ relative to $\omega$B97M-D4).

Figure~\ref{fig:28BNPAHs} collects the 28 BNPAHs
with finite direct SOC, showing their structures
together with the L-CC2 singlet-triplet gaps,
SOC magnitudes, oscillator strengths, and
root-mean-square (RMS) out-of-plane (OOP)
displacements. Molecular nonplanarity was
quantified as the RMS perpendicular displacement
of all non-hydrogen atoms from an orthogonal
least-squares plane. This global geometric
descriptor quantifies the extent of nonplanarity
rather than helicity itself.
Inspection of the molecular structures in Fig.~\ref{fig:28BNPAHs} suggests that the relationship between geometric helicity and direct ${\rm S}_1$-${\rm T}_1$ SOC depends on the nature of the helicene-like subunit rather than simply on the magnitude of the overall out-of-plane (OOP) distortion. 
For example, \textbf{BNPAH23} and \textbf{BNPAH27}, which contain pronounced helicene-like structural distortions, exhibit relatively large OOP displacements of 0.786 and 0.712~\AA, respectively, but comparatively weak SOC values of 0.022 and 0.065~cm$^{-1}$. 
In contrast, \textbf{BNPAH39} and \textbf{BNPAH45} exhibit smaller OOP displacements of 0.651 and 0.542~\AA, respectively, yet possess substantially larger SOC magnitudes of 0.436 and 0.437~cm$^{-1}$.
More strikingly, \textbf{BNPAH29} exhibits appreciable direct SOC (0.183~cm$^{-1}$) despite a relatively
small OOP displacement of 0.171~\AA.
These comparisons indicate that larger helicene-like distortions do not necessarily produce stronger SOC.
Instead, the local geometry and connectivity of the helicene-like subunit, together with the positions
of the boron and nitrogen atoms, influence the orbital mixing and electronic-state symmetry governing
singlet--triplet coupling. 
Thus, the observed SOC enhancement appears to depend on the specific molecular topology and pattern of structural distortion, rather than on global nonplanarity alone.
Furthermore, \textbf{BNPAH13} and \textbf{BNPAH45} possess nearly identical OOP displacements (0.549 and 0.542~\AA), whereas their SOC magnitudes differ by approximately a factor of 14, further highlighting the importance of molecular connectivity and local distortion patterns.

To investigate whether differences in the electronic character of the excited states contribute to the observed variation in SOC, NTOs were analyzed for five representative BNPAHs (\textbf{BNPAH13},  \textbf{BNPAH23}, \textbf{BNPAH27}, \textbf{BNPAH39}, and \textbf{BNPAH45}; Figure~S5 of the SI).
The leading NTO eigenvalues lie between 0.965 and 0.970 for ${\rm S}_1$ and between 0.882 and 0.970 for ${\rm T}_1$, with broadly similar hole and electron distributions for the two states of each molecule.
Despite this similarity, the SOC magnitudes differ substantially, indicating that neither the dominant NTO-pair character nor global nonplanarity alone accounts for the variation in coupling.

Helical or twisted geometries may relax symmetry restrictions and modify orbital mixing, as suggested by studies of azahelicenes and thiohelicenes.\cite{schmidt2007intersystem,bergwinkl2024enhanced} Nevertheless, the differing SOC values at similar OOP displacement indicate that the distortion pattern, electronic-state symmetry, and heteroatom connectivity also matter. 
We therefore regard helicity-associated enhancement as a plausible structural interpretation, rather than an established causal mechanism in the studied molecules.
The appreciable direct SOC of \textbf{BNPAH45}, highlighted in Fig.~\ref{fig:mechanism}, is consistent with its identification as one of the most promising candidates for direct rISC.

\begin{table}[!htbp]
\caption{Ten INVEST-BNPAHs with the highest estimated direct ${\rm T}_1\rightarrow {\rm S}_1$ rISC rates. 
Gaps are from L-CC2; SOC magnitudes are from TDDFT. 
The Marcus model uses $\lambda=0.20$~eV and $T=300$~K.}
\label{tab:direct_rates}
\begin{ruledtabular}
\begin{tabular}{lccc}
Molecule & $\Delta E_{\rm ST}$ (eV) & SOC (cm$^{-1}$) & $k_{\rm rISC}^{d}$ (s$^{-1}$) \\
\hline
\textbf{BNPAH39} & $-0.019$ & 0.436 & $7.47\times10^{6}$ \\
\textbf{BNPAH45} & $-0.017$ & 0.437 & $7.25\times10^{6}$ \\
\textbf{BNPAH40} & $-0.019$ & 0.284 & $3.16\times10^{6}$ \\
\textbf{BNPAH29} & $-0.028$ & 0.183 & $1.54\times10^{6}$ \\
\textbf{BNPAH28} & $-0.028$ & 0.117 & $6.29\times10^{5}$ \\
\textbf{BNPAH33} & $-0.023$ & 0.109 & $4.98\times10^{5}$ \\
\textbf{BNPAH36} & $-0.022$ & 0.085 & $3.03\times10^{5}$ \\
\textbf{BNPAH34} & $-0.023$ & 0.073 & $2.24\times10^{5}$ \\
\textbf{BNPAH27} & $-0.029$ & 0.065 & $1.96\times10^{5}$ \\
\textbf{BNPAH30} & $-0.027$ & 0.057 & $1.49\times10^{5}$ \\
\end{tabular}
\end{ruledtabular}
\end{table}

To assess the joint effects of electronic coupling and singlet-triplet energetics, the direct ${\rm T}_1\rightarrow {\rm S}_1$ rate was estimated within Fermi's golden-rule description,
\begin{equation}
 k_{\mathrm{rISC}}^{d}(T)=\frac{2\pi}{\hbar}\,\overline{|V^{d}|^2}\,F_{\mathrm{FC}}(T),
 \label{eq:golden}
\end{equation}
where $F_{\mathrm{FC}}(T)$ is the Franck--Condon-weighted density of states. With a classical Marcus form and three equally populated triplet sublevels, this becomes
\begin{equation}
 k_{\mathrm{rISC}}^{d}(T)=
 \frac{2\pi |V^{d}|^2}{3\hbar\sqrt{4\pi\lambda k_{\mathrm B}T}}
\exp\!\left[-\frac{(\Delta E_{\mathrm{ST}}+\lambda)^2}
 {4\lambda k_{\mathrm B}T}\right],
\label{eq:marcus}
\end{equation}
where $|V^{d}|$ denotes the spin-summed SOC norm, $\lambda$ the reorganization energy, and $T$ the temperature. 
The factor of $1/3$ averages over equally populated triplet spin microstates. 
We use L-CC2 vertical gaps as approximate driving forces and TDDFT couplings, together with $\lambda=0.20$~eV and $T=300$~K. 
The rates depend quadratically on SOC and exponentially on the assumed energy gap and reorganization energy.

Table~\ref{tab:direct_rates} lists the ten BNPAHs with the largest estimated rates. 
Across all 28 BNPAHs with finite SOC, the estimated direct-rISC rates span approximately $3.9\times10^3$ to $7.5\times10^6$~s$^{-1}$. 
The largest values occur for \textbf{BNPAH39} ($7.47\times10^6$~s$^{-1}$) and \textbf{BNPAH45} ($7.25\times10^6$~s$^{-1}$), followed by \textbf{BNPAH40} ($3.16\times10^6$~s$^{-1}$) and \textbf{BNPAH29} ($1.54\times10^6$~s$^{-1}$). 
These candidates combine relatively substantial coupling with small negative L-CC2 gaps. 
For the leading candidates, the rate ordering closely follows SOC because their gaps are similar.

A sensitivity analysis using $\lambda=0.10$ and
0.30~eV is provided in Tables~S1 and S2 of the SI. 
Rates decrease as $\lambda$ increases, 
but \textbf{BNPAH39} and \textbf{BNPAH45} retain the highest estimated rates. 
These values are not quantitative predictions of DF kinetics, as vertical gaps approximate free-energy differences, reorganization energies are assumed identical, and vibronic/environmental effects are omitted. 
Furthermore, the TDDFT description used for SOC does not reproduce the inverted ordering obtained with L-CC2, and the model omits indirect higher-state pathways. 
The results therefore provide a controlled, mixed-level screening of the 
direct channel rather than a calculation of total rISC or device-level triplet harvesting.

Our results reveal a strong dependence of direct ${\rm S}_1$--${\rm T}_1$ SOC on molecular framework within the INVEST class. 
In BNPAHs, intrinsic helicity emerges as a structural feature associated with enhanced SOC, although geometric distortion alone does not determine its magnitude. 
The favorable direct-rISC estimates for \textbf{BNPAH39} and \textbf{BNPAH45} identify promising candidates for further kinetic investigations. 
Importantly, the 46 BNPAHs examined here represent only a small fraction of the 30\,797 B,N-substituted PAHs, highlighting the considerable chemical space that remains unexplored. These findings motivate broader, systematic investigations of SOC alongside singlet--triplet inversion to guide the discovery of efficient INVEST emitters.

\section{Supplementary Information}
Figures S1--S3 depict INVEST systems with negative singlet--triplet gaps, 
Figure~S4 compares SOC values across DFT methods, and Figure~S5 presents representative NTOs. 
Tables S1 and S2 report Marcus-model direct-rISC rates for 28 BNPAHs at $\lambda = 0.10$ and $0.30$ eV, respectively. 
The INVEST85 dataset~\cite{INVEST85} provides the optimized geometries and L-CC2 excitation energies of all systems studied.

\section{Data Availability}
The data supporting the findings of this study are provided in the article and its supplementary material.

\section{Acknowledgments}
We acknowledge the support of the 
Department of Atomic Energy, Government
of India, under Project Identification No.~RTI~4007. 
All calculations have been performed using the Helios computer cluster, 
which is an integral part of the MolDis 
Big Data facility, 
TIFR Hyderabad \href{http://moldis.tifrh.res.in}{(http://moldis.tifrh.res.in)}.

\section{Author Declarations}

\subsection{Author contributions}
\noindent 
{\bf AM}: 
Conceptualization (equal); 
Analysis (equal); 
Data collection (equal); 
Review \& editing (equal);
Writing (equal).
{\bf RR}: Conceptualization (equal); 
Analysis (equal); 
Data collection (equal); 
Funding acquisition; 
Project administration and supervision; 
Resources; 
Review \& editing (equal);
Writing (equal).

\subsection{Conflicts of Interest}
The authors have no conflicts of interest to disclose.

\section{References}
\bibliography{ref} 
\end{document}